\documentclass[%
superscriptaddress,
preprint,
 amsmath,amssymb,
 aps,
pre,
]{revtex4-2}
\usepackage{amsmath}
\usepackage{amssymb}
\usepackage{graphicx} 
\usepackage{charter}
\usepackage[dvipsnames]{xcolor}
\usepackage{cleveref}
\usepackage[mathlines]{lineno}

\newcommand{\ddt}[1]{\frac{\mathrm{d}#1}{\mathrm{d}t}}
\newcommand{\mean}[1]{\left \langle #1 \right \rangle}
\newcommand{\ninv}{\frac{1}{N}}
\newcommand{\reaction}[3]{#1 & \;\xrightarrow{#3}\; #2}

\begin{document}
\title{Flocking by stopping -- \\ a novel mechanism of emergent order in collective movement}
\author{Yogesh Kumar KC}
\affiliation{School of Mathematics, Indian Institute of Science Education and Research, Thiruvananthapuram, Kerala, India, 695551.}

\author{Arshed Nabeel}
\email{Corresponding author: arshed@iisc.ac.in}
\affiliation{IISc Mathematics Initiative, Indian Institute of Science, Bengaluru, Karnataka, India, 560012.}
\affiliation{Center for Ecological Sciences, Indian Institute of Science, Bengaluru, Karnataka, India, 560012.}

\author{Srikanth Iyer}
\affiliation{Department of Mathematics, Indian Institute of Science, Bengaluru, Karnataka, India, 560012.}

\author{Vishwesha Guttal}
\affiliation{Center for Ecological Sciences, Indian Institute of Science, Bengaluru, Karnataka, India, 560012.}

\begin{abstract}
    Collective movement is observed widely in nature, where individuals interact locally to produce globally ordered, coherent motion. In typical models of collective motion, each individual takes the average direction of multiple neighbors, resulting in ordered movement. In small flocks, noise-induced order can also emerge with individuals copying only a randomly chosen single neighbor at a time. We propose a new model of collective movement, inspired by how real animals move, where individuals can move in two directions or remain stationary. We demonstrate that when individuals interact with a single neighbor through a novel form of halting interaction---where an individual may stop upon encountering an oppositely moving neighbor rather than instantly aligning---persistent collective order can emerge even in large populations. This represents a fundamentally different mechanism from conventional averaging-based or noise-induced ordering. 
Using deterministic and stochastic mean-field approximations, we characterize the conditions under which such 'flocking by stopping' behavior can occur, and confirm the mean-field predictions using individual-based simulations. Our results highlight how incorporating a stopped state and halting interactions can generate new routes to order in collective movement.
\end{abstract}

\maketitle
\section{Introduction}
    

Collective movement is a ubiquitous phenomenon in nature, where individuals move together, interacting with one another, resulting in emergent large-scale synchronization and pattern formation. Such emergent collective motion is observed in a wide range of systems across scales, ranging from cells and microorganisms, insects, fish, birds, large mammals, human crowds, etc.~\cite{alert_physical_2020, dyson_onset_2015, attanasi_collective_2014, jhawar_noise-induced_2020, cavagna_bird_2014, helbing_social_1995, bacik_lane_2023}. It is also observed in non-biological and engineered systems such as active colloids, Janus particles, microswimmers, etc.~\cite{ramaswamy_mechanics_2010, marchetti_hydrodynamics_2013}. As such, understanding the dynamics of collective movement is an active area of research that has attracted researchers across disciplines.

In particular, the spontaneous emergence of synchronized motion in a group of distinct individuals has been of special interest. The predominant paradigm in the field is to model the individuals as \emph{self propelled particles} that align their directions of motion to their neighbors, which can lead to coherent and synchronization of movement across the whole group~\cite{okubo_dynamical_1986, huth_simulation_1992, vicsek_novel_1995, chate_modeling_2008, couzin_collective_2002, escobedo_data-driven_2020, romanczuk_mean-field_2012, grosmann_active_2012,jadhav2022randomness}. Recently, alternative mechanisms have also been suggested through which collective order can emerge~\cite{das_flocking_2024, caprini_flocking_2023}.

Even in flocking systems with alignment interactions, there are different mechanisms through which ordered motion can arise. Consider the familiar example of the classic Vicsek model, where each individual aligns its direction of motion with all the neighbors within a finite radius. These alignment interactions---when noise fluctuations are low enough---results in highly synchronized movement across the whole group, even for very large group sizes~\cite{toner_long-range_1995}. This emergent ordering can occur even if the individuals interact with as few as two (instead of all) neighbors, chosen either based on spatial proximity or randomly from the whole group~\cite{dyson_onset_2015, chatterjee_three-body_2019, nabeel_data-driven_2023}. A fundamentally different mechanism of ordering takes place when individuals interact only with a single neighbor (chosen based on spatial proximity or randomly). In this case, order can still emerge, but only in finite group sizes~\cite{biancalani_noise-induced_2014, jhawar_noise-induced_2020, jhawar2020phila}. The emergent collective order here is called \emph{noise induced order}~\cite{horsthemke_noise-induced_1984}, and represents an entirely new mechanism of ordering, which disappears when the number of individuals grow to the thermodynamic limit.

In the real world biological collectives such as animal groups, individuals can move with variable speed and can even occasionally stop~\cite{puy_selective_2024, bazazi_intermittent_2012, ariel_individual_2014}. However, models of collective movement often assume constant-speed motion (but see \cite{mishra_collective_2012, hemelrijk_selforganized_2008,  klamser_impact_2021}). In this context, we present a simple, one-dimensional model of collective movement, where individuals can move in one of two directions, or remain stationary. In this model, we show that collective order can emerge and persist for large group sizes, even when individuals interact with a single neighbor. This is in contrast with previous (constant speed) models, where the emergent order with single neighbor interactions is noise-induced and disappears when the group size grows~\cite{biancalani_noise-induced_2014, jhawar_noise-induced_2020}. 

This new mechanism for emergent order is a surprising consequence of a new type of interaction which we call \emph{halting interactions}. Here, an individual can stop moving when encountering a neighbor moving in a different direction, instead of immediately copying its direction. 
We use mean-field descriptions of the model dynamics in the form of ordinary and stochastic differential equations to characterize conditions in which ordered motion can emerge. The mean field predictions match well with individual-based stochastic simulations of the model dynamics. In summary, our work demonstrates how speed variability can offer novel routes to order in collective movement.

\section{A simple model for variable-speed collective movement} \label{sec:model}

\begin{figure*}
    \centering
    \includegraphics[width=\linewidth]{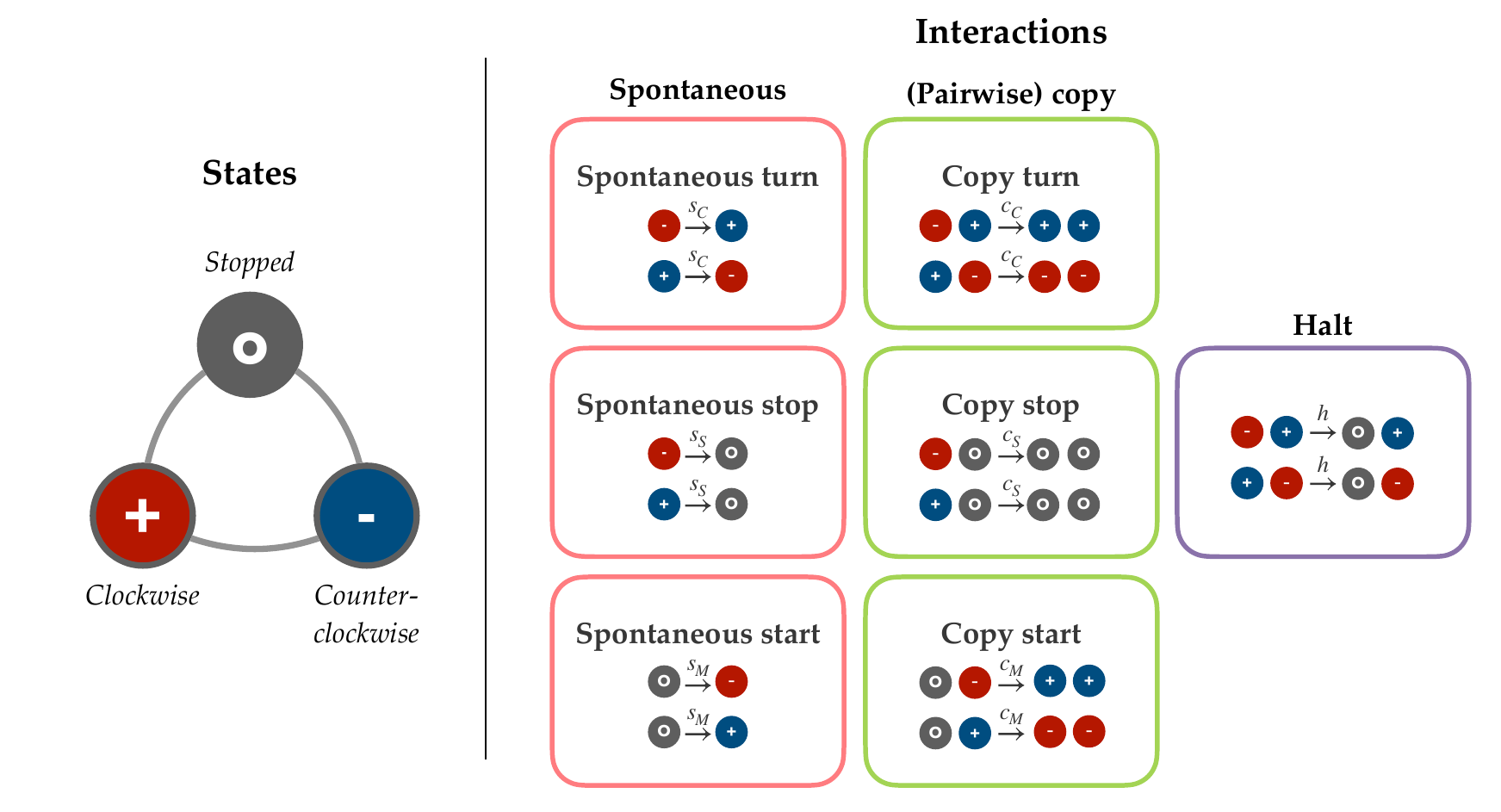}
    \caption{\textbf{A schematic illustration of the model.} \textit{Left:} Each individuals can be in one of three states + (moving clockwise), - (moving counterclockwise), or 0 (stopped). \textit{Right:} The individuals can switch between these states either spontaneously or by interacting with a random neighbor, through the different types of interactions as shown.}
    \label{fig:schematics}
\end{figure*}

In real world animal collectives, individuals typically move with variable speed, and often exhibit stop-and-go movement~\cite{klamser_impact_2021, kent_speed-mediated_2019, bazazi_intermittent_2012, ariel_individual_2014}. However, most models of collective movement ignore this aspect of movement. Inspired by the experiments involving locusts, fish etc. moving in annular arenas~\cite{buhl_disorder_2006, jiang_identifying_2017}, where individuals can effectively move only along clockwise and counter-clockwise movement, we present the following simplified model of variable speed: We consider $N$ individuals where each individual can take one of three states; $X_+$ (moving clockwise), $X_-$ (moving counterclockwise) and $X_0$ (stopped). Individuals can change their state of motion either spontaneously, or by interacting with random individual from the group, as described below and summarized in \cref{fig:schematics}. Our assumption that discrete states can represent movement variables is justified in many experimental contexts where animals perform motion in effective 1-dimensional systems such as an annular arena with relatively thin annual region (e.g. ~\cite{buhl_disorder_2006, jiang_identifying_2017}).

\paragraph*{Spontaneous switching.} First, individuals can spontaneously change their movement state to a new state with some specified rate. We assume that spontaneous movement initiations, spontaneous stops and spontaneous direction changes occur at rates $s_M, s_S$ and $s_C$ respectively.

\begin{align}
    \reaction{X_0}{X_+}{s_M}    \label{reac:1} \\
    \reaction{X_0}{X_-}{s_M}    \\
    \reaction{X_+}{X_0}{s_S}    \\
    \reaction{X_-}{X_0}{s_S}    \\
    \reaction{X_+}{X_-}{s_C}    \\
    \reaction{X_-}{X_+}{s_C}    
\end{align}

\paragraph*{Copy interactions.} Second, an individual can choose a random neighbor from the group and copy its motion state. We assume different rates $c_M, c_S$ and $c_C$ for copy events that results in movement initiation, movement stop and direction change, respectively. In the equations below, one may take the first individual as the focal individual and the second as the randomly chosen neighbor from the group.

\begin{align}
    \reaction{X_0 + X_+}{X_+ + X_+}{c_M}    \label{eq:cm1} \\
    \reaction{X_0 + X_-}{X_- + X_-}{c_M}    \label{eq:cm2} \\
    \reaction{X_+ + X_0}{X_0 + X_0}{c_S}    \\
    \reaction{X_- + X_0}{X_0 + X_0}{c_S}    \\
    \reaction{X_- + X_+}{X_+ + X_+}{c_C}    \\
    \reaction{X_+ + X_-}{X_- + X_-}{c_C}    
\end{align}

\paragraph*{Halting interactions.} In the copying interactions described above, individuals switched their directions of motion when they encountered an individual moving in the opposite direction. In such cases, alternatively, the individual may first reduce the speed, may even come to a stop and then change the direction. We model this scenario via a \textit{`halting interaction'}, where the focal individual stops the movement as a consequence of interaction with the neighbor who is moving in the opposite direction as itself. The corresponding equations are:

\begin{align}
    \reaction{X_+ + X_-}{X_0 + X_-}{h}    \\
    \reaction{X_- + X_+}{X_0 + X_+}{h}  \label{reac:n}
\end{align}


The dynamics of this system is governed by the 7 rate parameters. Our main goal in the remainder of this manuscript is to characterize the different dynamical regimes of the system in terms of these rate parameters. We note that the class of \emph{constant speed} pairwise interaction models previously considered in literature~\cite{biancalani_noise-induced_2014, jhawar_deriving_2019}, where individuals only change direction and do not stop, arises as a special case of our model, with $c_M, c_S, s_M, s_C, h =0$, and the system initialized with no individuals in a stopped state.

\section{Model dynamics: flocking by stopping}
\subsection{Mean field dynamics: stopping promotes collective order}
\begin{figure*}
    \centering
    \includegraphics[width=.95\linewidth]{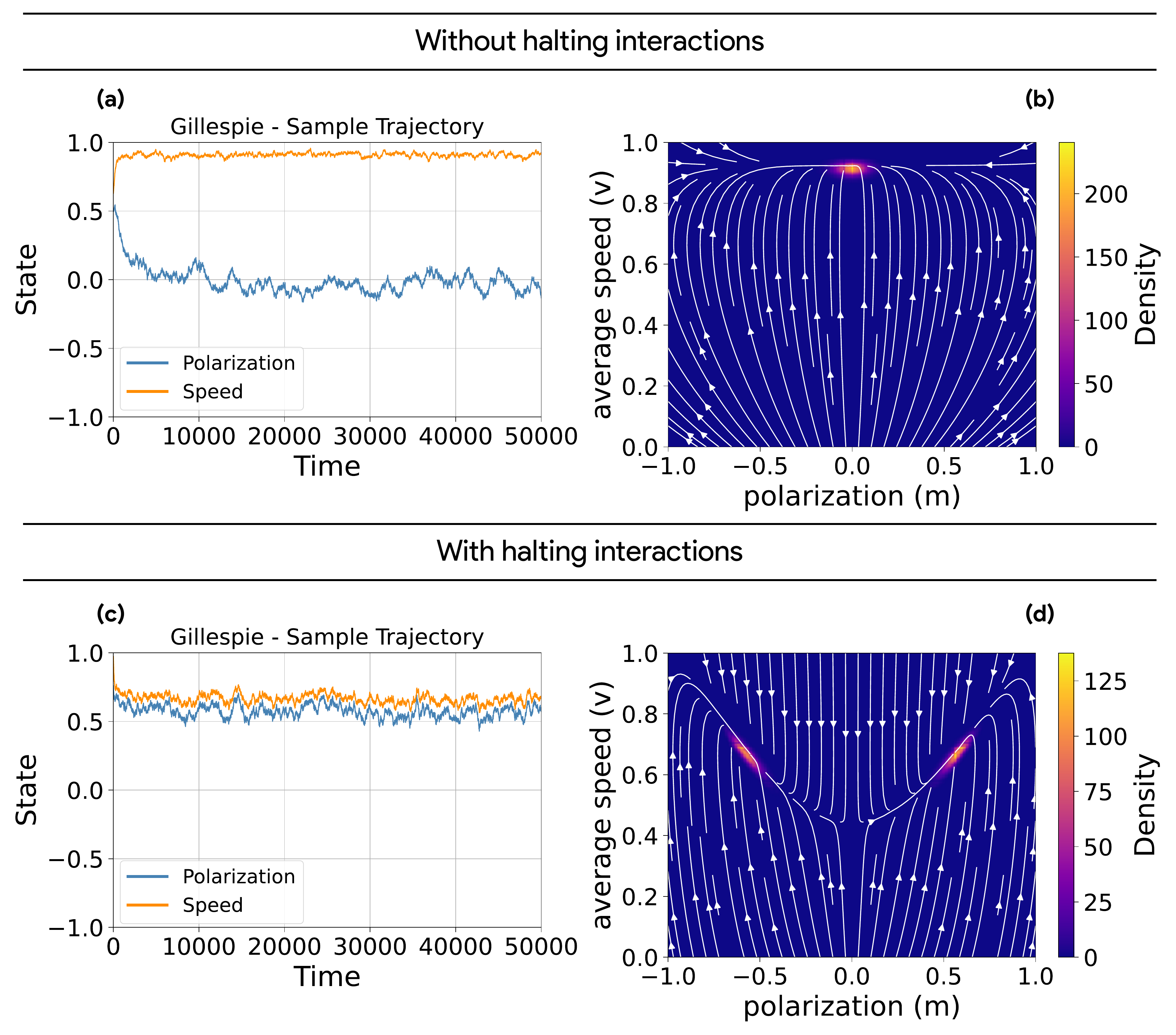}
    \caption{\textbf{Collective order via stopped state and halting interactions.} \textit{Top:} Without halting interactions ($h = 0$), the disordered state is the stable-state. \textit{Bottom:} With halting interactions ($h = 7$), the system can show ordered dynamics. \textit{(a) and (c):} Example trajectories of order parameters $m, v$, with and without halting interactions respectively. \textit{(b) and (d):} Phase plane of the mean-field model with the background heatmap showing steady state histograms, computed from Gillespie simulations for $N=500$, with and without halting interactions respectively.  (Parameter values: $s_S = s_M = s_C = c_S = c_C = 0.2, h \in \{0,7\}$ and $c_M = 2$.)}
    \label{fig:flocking}
\end{figure*}

We denote as $N_+(t), N_-(t)$ and $N_0(t)$ the number of individuals at time $t$ in the $X_+, X_-$ and $X_0$ states, respectively. We then define the fraction of individuals in each state as $x_+ = N_+/N, x_- = N_-/N, x_0 = N_0/N$, where $N = N_+ + N_- + N_0$ is the total number of individuals. Since $x_+ + x_- +x_0 = 1$, the state of the system is fully described by any two of these variables.

As an alternative characterization of the state of the system, we define two \emph{order parameters}, $m = x_+ - x_-$ and $v = x_+ + x_-$. The parameter $m$ can be thought of as the degree of alignment in the system; here $m = 1$ (corresponding to all individuals moving in the counter-clockwise direction) and $m = -1$ (when all individuals are moving in the clockwise direction) both represent high degree of alignment among individuals, whereas $m \approx 0$ represents an incoherent motion with the group nearly equally split between two directions. Similarly, $v$ represents the total fraction of moving individuals, and can be thought of as the \emph{average group speed}, with $v=1$ when all individuals are moving.

One can describe the deterministic mean field dynamics using coupled ordinary differential equations describing the time-evolution of $m$ and $v$. Using the \textit{chemical Langevin method}~\cite{gillespie_chemical_2000, jhawar_deriving_2019}, the following mean field ODEs can be derived from the interaction model (see Appendix for the full derivation):

\begin{align}
    \ddt{m} &= (a_1 - a_2v)m \label{eq:ode-m} \\
    \ddt{v} &= b_1 + b_2 m^2 + b_3 v - b_4 v^2 \label{eq:ode-v}
\end{align}

with
\begin{align*}
a_{1} &= -(s_{M}+2s_{C}) + (c_{M} - c_{S}), \\
a_{2} &= c_{M} - c_{S}, \\
b_{1} &= 2s_{M}, \\
b_{2} &= \frac{h}{2}, \\
b_{3} &= -(2s_{M} + s_{S}) + (c_{M} + c_{S}) \text{, and} \\
b_{4} &= (c_{M} + c_{S}) + \frac{h}{2}. \\
\end{align*}

Our system admits the following equilibrium solutions;
\begin{align*}
    m_1^* &= 0, & v_1^*&=\frac{b_3 + \sqrt{b_3^2 + 4 b_1 b_4}}{2b_4},  \\
    m_2^* &= + \sqrt{\frac{b_4v_2^{*2} - b_3 v_2^* - b_1}{b_2}}, & v_2^* &= \frac{a_1}{a_2} \quad \text{and} \\
     m_3^* &= - \sqrt{\frac{b_4v_3^{*2} - b_3 v_3^* - b_1}{b_2}}, & v_3^* &= \frac{a_1}{a_2}
\end{align*}
When $a_1/a_2 < \left(b_3 + \sqrt{b_3^2 + 4 b_1 b_4}\right)/{2b_4}$, the disordered ($m = 0$) state is stable, and when $a_1/a_2 > \left(b_3 + \sqrt{b_3^2 + 4 b_1 b_4}\right)/{2b_4}$, the ordered ($m \neq 0$) state is stable. In other words, there exist parameter regimes for the rates such that our system admits an ordered state as a mean-field stable solution, where the individuals consistently move in the clockwise or counterclockwise direction with non-zero polarization and speed. Put simply, this means that the individuals can display ordered, `flocking' motion, even with simple pairwise interactions. This is in contrast with typical `constant speed' models of collective movement, which requires a ternary (or higher order) interactions to exhibit ordered motion in the mean-field limit~\cite{biancalani_noise-induced_2014,jhawar_deriving_2019, nabeel_data-driven_2023}. Furthermore, the ordered state exists only when $b_2 = h/2 > 0$. In other words, the halting interactions are crucial in the model to create ordered collective motion.

This is illustrated in \cref{fig:flocking}. Without halting interaction ($h=0$), the disordered state ($m=0$) is the only one stable state (\cref{fig:flocking} a, b). In contrast, when halting interactions are present, i.e. when $h>0$, we see that the system shows order (\cref{fig:flocking} c, d). The phase plots in panels (b) and (c) show the dynamics according to the mean-field ODE approximations. We also see that the steady-state histograms in these panels, obtained from Gillespie simulations of the microscopic interaction model, closely matches the mean field predictions, reaffirming that the mean-field ODEs are a good description of the behavior of the system for large group sizes.

How does the the presence of additional stopped state $X_0$ and halting interactions break symmetry between the $X_+$ and $X_-$ state and lead to emergent order? To understand this, first note that state changes through spontaneous interactions (eqq. 1-6) are symmetric between $X_+$ and $X_-$ and hence cannot lead to symmetry-breaking. Similarly, state-changes due to copy interactions between $X_+$ and $X_-$ states are also symmetric, which is why in a two-state system (without the stopped state) copy interactions cannot create deterministic order. Whereas in our model, the net flow into the $X_+$ state through pairwise interactions is $(c_M - c_S)x_+x_0 - hx_+x_-$, and the flow into $X_-$ state is $(c_M - c_S)x_-x_0 - hx_+x_-$. These flows are asymmetric: if for instance $x_+ > x_-$, the flow into $X_+$ will also be greater (assuming $c_M > c_S$), establishing a feedback loop and breaking symmetry. Therefore, the presence of the stopped state $X_0$ helps to create order by breaking the symmetry between the flows into the $X_+$ and $X_-$ state. 

To understand the role of the halting interactions, first note that $x_0$ being nonzero is essential for the symmetry-breaking argument presented above. When $x_0 = 0$, the symmetrizing mechanism due to pairwise copy interactions between $X_+$ and $X_-$ takes hold and destroys order. Therefore, somewhat counterintuitively, a mechanism that ensures a steady flow from the moving states into the stopped $X_0$ state is essential for the maintenance of order. Now, consider the net flow into the $X_0$ state, which is $-2s_M x_0 + s_S(x_++x_-) + (c_S - c_M)(x_+ + x_-)x_0 + hx_+x_-$. When $x_0$ is equal to or close to $0$, the flow into $X_0$ is $s_S(x_+ + x_-) + hx_+x_-$, since other terms go to 0. Out of these, the flow due to spontaneous stopping, $s_S(x_+ + x_-)$, again has a symmetrizing effect between $X_+$ and $X_-$. Thus only the $hx_+x_-$ term provides this steady flow into $X_0$ without decreasing order as a side effect. Thus, we see that both the $X_0$ state and halting interactions are crucial for this new mechanism of symmetry breaking.

To summarize, the stopped state and halting interactions allow the system to achieve ordered motion even in large groups, with simple pairwise interactions. This is in contrast with constant-speed models, were higher-order interactions are typically necessary for order to emerge~\cite{dyson_onset_2015,jhawar_deriving_2019}.


\subsection{Role of copy interactions}
\begin{figure*}
    \centering
    \includegraphics[width=1\linewidth]{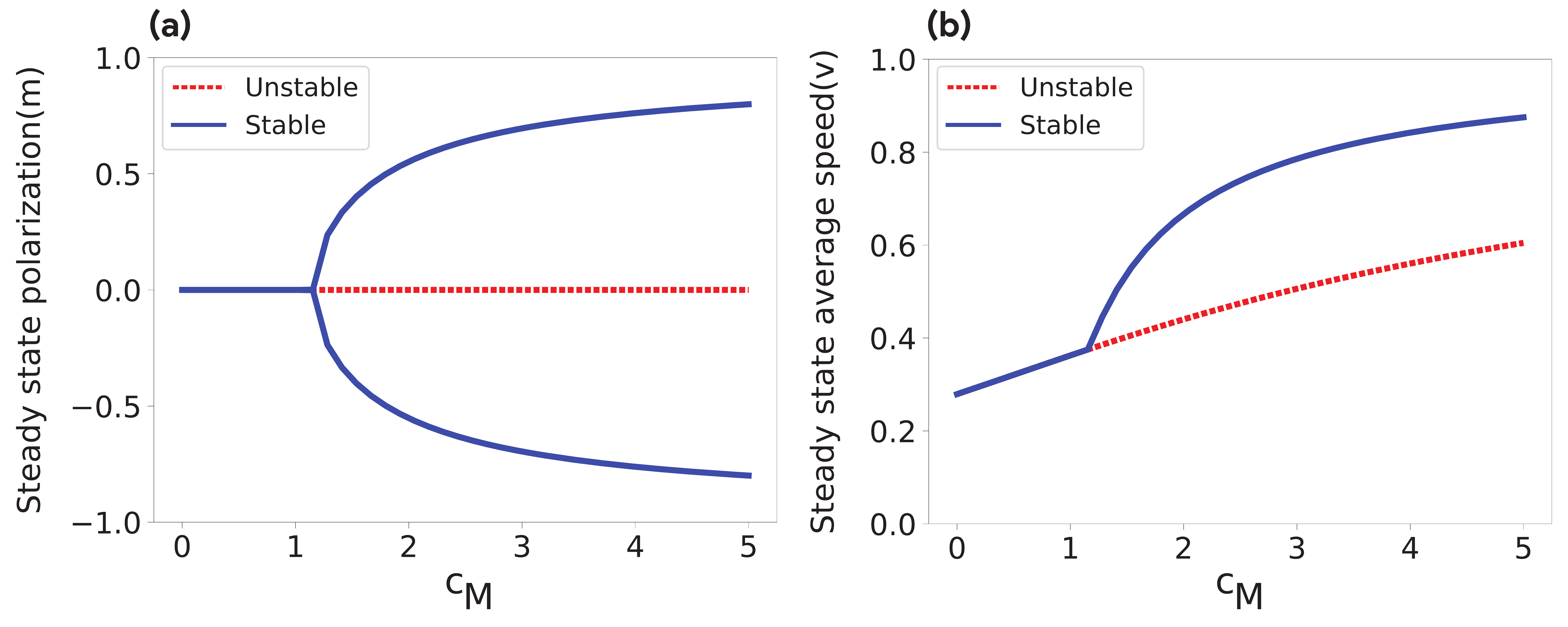}
    \caption{\textbf{Role of copy interactions in collective order.} The ordered state is a deterministic stable state for only sufficiently high values of $c_M$. \textit{(a):} As $c_M$ increases, the order $m$ undergoes a bifurcation. \textit{(b):} A similar bifurcation occurs for speed $v$.(Parameter values: $s_S = s_M = s_C = c_S = c_C = 0.2, h = 7$ and $c_M \in [0,5]$)}
    \label{fig:bifurcation}
\end{figure*}

Even in the presence of halting interactions, ordered movement can emerge only when $c_M$, the rate at which stopped individuals copy moving individuals, is sufficiently high. This is illustrated in the bifurcation diagrams in \cref{fig:bifurcation}, with $c_M$ as the control parameter. When $c_M$ is low, only the disordered state is stable. Once $c_M$ crosses the bifurcation threshold, the ordered state becomes stable. Beyond this threshold, the magnitude of $m$ increases with $c_M$. In other words, as $c_M$ increases, the system forms increasingly ordered flocks. The average speed (or equivalently, the fraction of moving individuals) $v$ also shows a bifurcation behavior, although $v$ is non-zero even for $c_M$ below the threshold value.

\subsection{Stochastic dynamics and finite-size effects}

So far, we analyzed the dynamics of the system in the large-$N$ limit, with mean-field ODE approximations. When the number of individuals in the flock is small, the effect of individual fluctuations on the group dynamics cannot be ignored, and can have non-trivial, \textit{noise-induced} effects on the flocking dynamics~\cite{biancalani_noise-induced_2014,jhawar_noise-induced_2020}. The stochastic dynamics of finite flocks can be approximated with a stochastic differential equation (SDE). Deferring the full derivation to appendix \ref{sec:ax-sde} as before, the coupled {\it It\^{o}} SDEs corresponding to our model is given by,

\begin{align}
    \ddt{m} &= (a_1 - a_2v)m + \sqrt{\ninv \left(f_1 + f_2 m^2 +f_3 v + f_4 v^2\right)} \; \eta_m(t) \label{eq:sde-m} \\
    \ddt{v} &= b_1 + b_2 m^2 + b_3 v - b_4 v^2 + \sqrt{\ninv \left( e_1 + e_2 m^2 + e_3 v + e_4 v^2\right)} \; \eta_v(t) \label{eq:sde-v}
\end{align}
where the coefficients $a_i, b_i, f_i, e_i$ are, as before, functions of the rate parameters of the model (see appendix ~\ref{sec:ax-sde} for details). $\eta_m$ and $\eta_v$ are uncorrelated noise fluctuations with $\mean{\eta_m} = \mean{\eta_v} = 0$, $\mean{\eta_m(t) \eta_m(t')} = \mean{\eta_v(t) \eta_v(t')} = \delta(t-t')$ and $\mean{\eta_m(t) \eta_v(t)} = 0$. Note that the strength of the stochastic fluctuations fall as $1/\sqrt{N}$ as $N$ increases, and the model reduces to the ODEs in \cref{eq:ode-m,eq:ode-v} in the limit $N \to \infty$. However, for small values of $N$, multiplicative noise can have interesting effects on dynamics.

\begin{figure*}
    \centering
    \includegraphics[width=\linewidth]{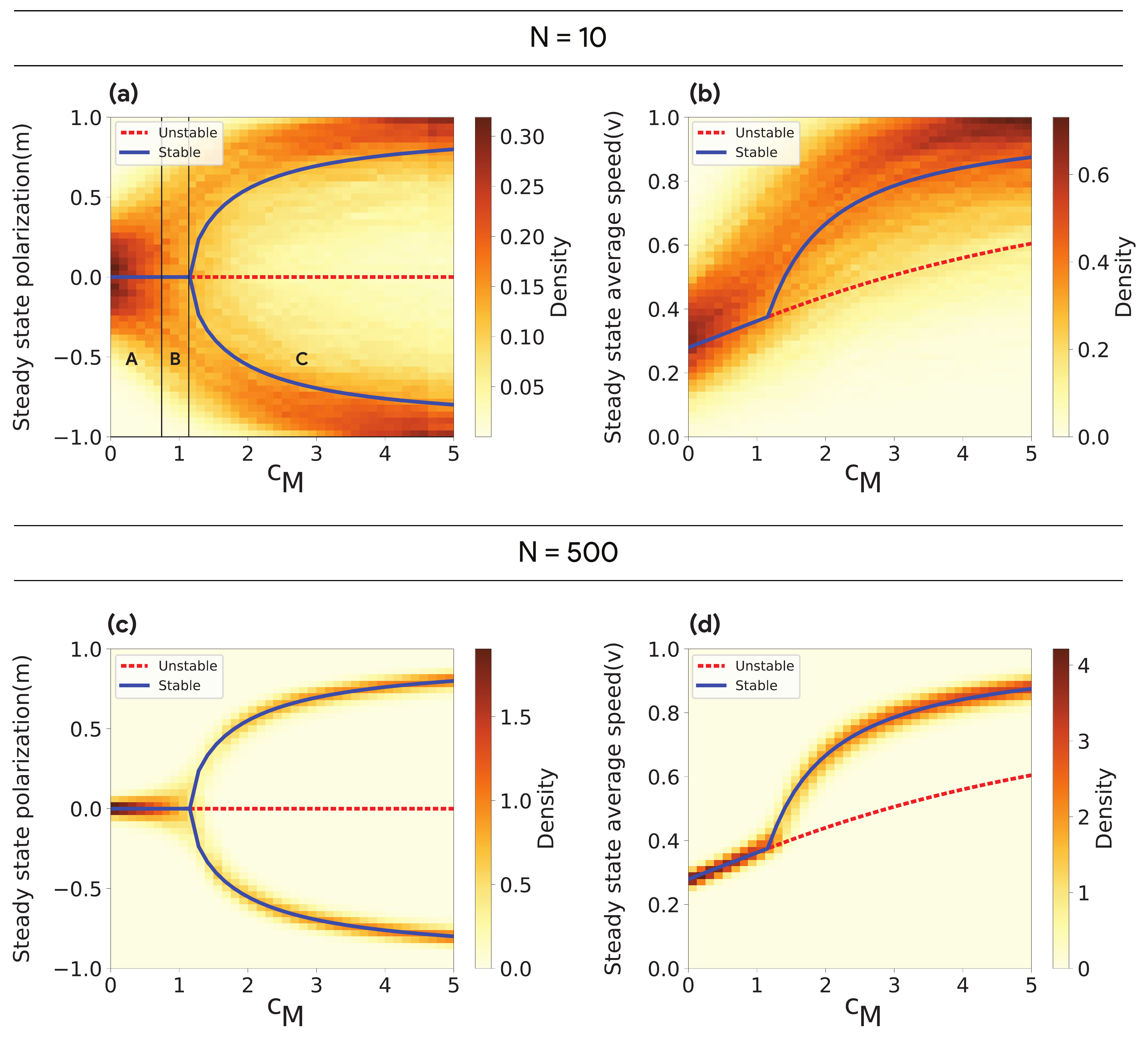}
    \caption{\textbf{Stochastic dynamics and noise effects in small flocks.} Histograms of $m$ and $v$ obtained from stochastic simulations of the SDE model, as a function of $c_M$, with $N=10$ (top row, panels (a) and (b)) and $N=100$ (bottom row, panels (c) and (d)). The bifurcation plots, representing the mean-field predictions, are overlaid on the normalized histograms (heatmaps). Top row: For $N=10$, we identify three ranges of copying interactions $c_M$, {\bf A, B} and {\bf C}. In region {\bf A} (representing the region of disorder), the histograms match the expectations of mean-field theory. However, in region {\bf B}, the histograms show a finite order, suggesting a noise-induced order, which deviates from the mean-field stable state of disorder. In the region {\bf C}, the obesrved order is quantitatively higher than the mean-field predictions, indicating a case of noise facilitating the order. Bottom row: For $N=100$, we do not observe the noise-induced or noise-facilitated order, consistent with mean-field theory.}
    \label{fig:stochastic}
\end{figure*}

The results of the It\^{o} SDE simulations using Euler-Mayurama numerical integration scheme are illustrated in~\Cref{fig:stochastic}, which shows the histograms of $m$ and $v$ in steady states overlaid on the mean-field bifurcation diagram. For large $N$ (panels c, d), the steady-state histograms from the stochastic simulations of SDEs closely match the mean-field predictions from the deterministic approximation. However, for small $N$, finite-size stochasticity starts to have non-trivial effects on the observed dynamics. Specifically, for any fixed value of $c_M$, we see that the observed net order $|m|$ and speed $v$ is higher in the stochastic simulations, than what is predicted by the mean-field ODEs (see region marked \textit{C} in panel (a)). In addition, the histogram bifurcates into a bi-modal histogram even before the critical threshold value of $c_M$ (see region marked \textit{B} in panel (a)), suggesting a noise-induced ordered state. In Appendix~\ref{sec:ax-sde-ssa}, we show further results to demonstrate that the SDE closely approximates the stochastic dynamics of the system for small as well as large group sizes.

In summary, we see that finite-size stochasticity can induce ordered movement in regimes where the deterministic ODEs predict no order (region {\bf B} in ~\Cref{fig:stochastic}(a)) and can enhance order in regions in which mean-field theory predicts an ordered stable state (region {\bf C} in ~\Cref{fig:stochastic}(a)) . Note that this mechanism is complementary to the case previously demonstrated in constant-speed models~\cite{biancalani_noise-induced_2014,jhawar_noise-induced_2020}. The mechanism of noise-induced order from constant speed models is also present in our system, and manifests in small group sizes when $h = 0$.

\section{Discussion}

In this paper, taking inspiration from real-world animal behavior, we propose a simple model of collective movement in which individuals can occasionally stop moving. The individuals can change their movement directions or movement state (stationary/moving) spontaneously or through pairwise interactions, interacting with one random neighbor at a time. We propose a new type of interaction, called a \emph{halting interaction}, where an individual stops moving when encountering an oppositely-moving neighbor, and show that halting interactions can lead to collectively ordered movement that persists even in very large group sizes. Previously, pairwise interactions have been shown to create \emph{noise-induced order} in small-sized flocks~\cite{biancalani_noise-induced_2014,jhawar_noise-induced_2020}. In contrast, our mechanism of collective order, which we call `flocking by stopping', is deterministically stable, and persists even in large group sizes: for individuals interacting through only pairwise interactions, this is a novel route to collective order.

The emergent order is a result of the interplay between the halting interactions (where individuals encountering neighbors moving in the opposite direction stop their motion) and `copy-start' interactions, where a stopped individual interacting with a moving neighbor starts moving in the neighbors direction. Consider an individual in state $X_+$ sequentially encountering two neighbors in state $X_-$, and interacting with them via a halting interaction followed by a copy-start interaction. This can be thought of as a ternary interaction (where individual encountering two neighbors moving in the opposite direction copies their direction) stretched out in time. Ternary interactions have previously been shown to create (deterministic) order in collective movement~\cite{jhawar_deriving_2019,dyson_onset_2015}. Note that without a the stopped state $X_0$, this effect is absent; the individual would copy the first neighbor and go to state $X_-$, and the second interaction would have no effect. Another interpretation is that of the focal individual accumulating information about multiple neighbors over two steps, and changing its direction only when it encounters two individuals moving in the opposite direction. Instead of using a latent variable to represent the accumulated information, the individual uses its own motion state: for instance, an individual moving clockwise stops moving when encountering the first neighbor moving anticlockwise, and changes direction if it encounters one more anticlockwise neighbor.

In the real world, most organisms move with varying speeds and can even stop; yet variable-speed models are relatively uncommon in literature (but see~\cite{mishra_collective_2012,klamser_impact_2021}). Our work suggests that speed variability, may be of significance in generating ordered collective movement. The proposed halting mechanism, corresponds to the tendency of individuals to slow down or stop when encountering misaligned neighbors. This is consistent with empirical observations that individuals tend to slow down when the local polarization around the individual is low~\cite{mishra_collective_2012}.

We have presented our model as a one-dimensional model, with three discrete states (moving clockwise, moving counterclockwise, and stopped). This model mirrors experimental settings with fish, locusts, etc., where the individuals move in an annular arena, and exhibits variable speed or even intermittent movement~\cite{buhl_disorder_2006, jiang_identifying_2017}. Is the broad mechanism of `flocking by stopping' valid in the case of more general two-dimensional movement with continuously variable speed? This is a question for future research through both experiments as well as models.

Our model has parallels to certain models of opinion dynamics, here individuals can be in an `uncommitted' state (mirroring the stopped or 0 state in our model), and can have \textit{cross-inhibition interactions}, similar to the halting interactions in our model. Cross inhibition models are known to create and promote group consensus in opinion dynamics~\cite{seeley_stop_2012, reina_design_2015, reina_cross-inhibition_2023}. However, our model has some crucial differences from these cross-inhibition models. First, our model was developed in the context of collective movement of animal groups, to explain mechanisms of the spontaneous emergence of order. Second, inspired by this context, our model includes more interactions, namely \textit{copy turn} and \textit{copy stop}, which reflect actual animal movement behavior: these interactions are absent in the classical cross-inhibition models~\cite{seeley_stop_2012, marshall_optimal_2009}. Finally, to make the connection to variable speed collective movement, we have introduced the new order parameter $v = x_+ + x_-$, which can be interpreted as the average individual speed.

In summary, we demonstrate a simple model of variable-speed collective movement with pairwise interactions, and demonstrate a new route to ordered collective movement in this system. The model suggests that speed variability, and the ability to slow down when encountering mis-aligned neighbors, can be of crucial importance in achieving collective order. We hope that our model acts as a framework for analyzing the collective movement of real-world animal groups, or for building more realistic models of collective movement with 2- or 3-dimensional motion, continuously variable speed, local interactions, etc. 

\section*{Acknowledgments}
We thank Andreagiovanni Reina for insightful discussions about the model, and about connections to cross inhibition models. VG acknowledges the Science Education and Research Board, Government of India, and AN acknowledges the Ministry of Education, Government of India for funding.

\section*{Data and Code Availability}
The code used in this study is available in the GitHub repository \url{https://github.com/tee-lab/pair_stop_paper}

\bibliography{references}

\appendix
\section{\label{sec:ax-sde} Deriving mean-field descriptions of the microscopic interaction model} 

In this section, we outline the steps involved in deriving the mean-field SDE descriptions for the dynamics of of the model described in section \ref{sec:model} and \cref{fig:schematics}. Recall that the state of the system can be described in terms of $x_-, x_0$ and $x_+$, which represent the fraction of individuals in states $X_-, X_0$ and $X_+$ respectively. We use Gillespie's \emph{Chemical Langevin} method to derive SDE approximations describing the time evolution of the state variables~\cite{gillespie_chemical_2000}. Subsequently, we can invoke It\^{o}'s lemma to write down the SDEs for the order parameters $m = x_+ - x_-$ and $v = x_+ + x_-$.

Our model involves 14 reactions through which the state variables can change. We start by writing down the \textit{propensity functions} and the \textit{state change matrix}. Let $\mathbf{x} = [x_- \quad x_0 \quad x_+]^T$ be the state vector. The propensity function $a_i(\mathbf{x})$ denotes the probability of reaction $i$ to occur, given the current state is $\mathbf{x}$. For example, for the reaction ${X_- + X_+} \xrightarrow{c_C}{X_+ + X_+}$, the propensity is given by $a_i(\mathbf{x}) = c_C x_+ x_-$. Let $\mathbf{a} = [a_1 \; \ldots \; a_{14}]$.

For the state change matrix $V$, the entry $v_{ij}$ corresponds to the change in number of individuals of state $j$, when reaction $i$ occurs. For example, the row in the state-change matrix corresponding to the above reaction is given by $[-1 \; 0 \; 1]$, since one individual changes from the $-$ state to the $+$ state.

Under mild assumptions, we can then write down an SDEs (interpreted according to It\^{o} conventions) for each $x_j$ (where $j \in \{-, 0, +\}$) as,

\begin{align}
    \frac{dx_j}{dt} &= \sum_{i=1}^{14} v_{ij} a_{ij} + \frac{1}{\sqrt N}\sum_{i=1}^{14} \sqrt{v_{ij} a_{ij}} \; \eta_i(t)
\end{align}

where the summations run over the 14 reactions, and $\eta_i(t)$ are independent white noise terms corresponding to each reaction. Writing out the equations and simplifying, we get the following SDEs for $x_-$ and $x_+$ respectively (recall that $x_+$ and $x_-$ describe the state of the system completely, since $x_+ + x_- + x_0 = 1$):




\begin{align}
    \frac{dx_+}{dt} &= s_{M}x_{0} - (s_{S} + s_{C})x_{+} +s_{C}x_{-} + (c_{M} - c_{S})x_{0}x_{+} - hx_{+}x_{-}  \\
    &+ \frac{1}{\sqrt{ N }}\sqrt{s_{M}x_{0} + (s_{S} + s_{C})x_{+} + s_{C}x_{-} + (c_{M} + c_{S})x_{0}x_{+} + 2(c_{C} + h)x_{-}x_{+}} \; \eta_{+}(t)   \\
    \frac{dx_-}{dt} &= s_{M}x_{0} + s_{C}x_{+} - (s_{S}+s_{C})x_{-} + (c_{M} - c_{S})x_{0}x_{-} - hx_{+}x_{-} \\
    &+ \frac{1}{\sqrt{ N }}\sqrt{ s_{M}x_{0} + (s_{S} + s_{C})x_{-} + s_{C}x_{+} + (c_{M} +c_{S})x_{0}x_{-} + 2(c_{C} + h)x_{-}x_{+} } \; \eta_{-}(t)
\end{align}

Using the relation $m = x_- - x_+$ and $v=x_- + x_+$, and some algebra, we can transform the above into the following It\^{o} SDEs for $m$ and $v$:

\begin{align}
    \ddt{m} &= (a_1 - a_2v)m + \sqrt{\ninv \left(f_1 + f_2 m^2 +f_3 v + f_4 v^2\right)} \; \eta_m(t) \label{eq:sde-m} \\
    \ddt{v} &= b_1 + b_2 m^2 + b_3 v - b_4 v^2 + \sqrt{\ninv \left( e_1 + e_2 m^2 + e_3 v + e_4 v^2\right)} \; \eta_v(t) \label{eq:sde-v}
\end{align}

where,
\begin{align*}
a_{1} &= -(s_{M}+2s_{C}) + (c_{M} - c_{S}) \\
a_{2} &= c_{M} - c_{S} \\
b_{1} &= 2s_{M} \\
b_{2} &= \frac{h}{2} \\
b_{3} &= -(2s_{M} + s_{S}) + (c_{M} + c_{S}) \\
b_{4} &= (c_{M} + c_{S}) +\frac{h}{2} \\
e_{1} &= 2s_{M} \\
e_{2} &= -\frac{h}{2} \\
e_{3} &= -2s_{M} + s_{S} + (c_{M} + c_{S}) \\
e_{4} &= \frac{h}{2} - (c_{M}  + c_{S}) \\
f_{1} &= 2s_{M}  \\
f_{2} &= - \frac{c_{C}}{2} - \frac{3}{4}h \\
f_{3} &= -2s_{M} + \frac{3}{2}s_{S} + s_{C} + (c_{M} + c_{S}) \\
f_{4} &= \frac{c_{C}}{2} + \frac{3}{4}h - (c_{M} + c_{S})
\end{align*}

The noise terms $\eta_m$ and $\eta_v$ are the noise fluctuations in $m$ and $v$ respectively. 
Since $\eta_m = \eta_+ - \eta_-$ and $\eta_v = \eta_+ + \eta_-$, if we assume that $\eta_+$ and $\eta_-$ are uncorrelated (an assumption that is validated by the simulations in Appendix~\ref{sec:ax-sde-ssa}), it follows that $\eta_m$ and $\eta_v$ also uncorrelated.
As $N \to \infty$, the stochastic term disappears due to the $1/\sqrt{N}$ dependence. This gives us the ODE approximation, valid for large group sizes, as:

\begin{align}
    \ddt{m} &= (a_1 - a_2v)m \\
    \ddt{v} &= b_1 + b_2 m^2 + b_3 v - b_4 v^2
\end{align}

\section{\label{sec:ax-sde-ssa} Stochastic dynamics of the model}

Strictly speaking, stochastic differential equations in \cref{eq:sde-m,eq:sde-v} are approximations of the actual microscopic dynamics of the microscopic model. In this section, we demonstrate that the SDE approximation closely matches the true dynamics of the microscopic model, across parameter ranges.

\begin{figure*}
    \centering
    \vspace{-20pt}
    \includegraphics[width=0.75\linewidth]{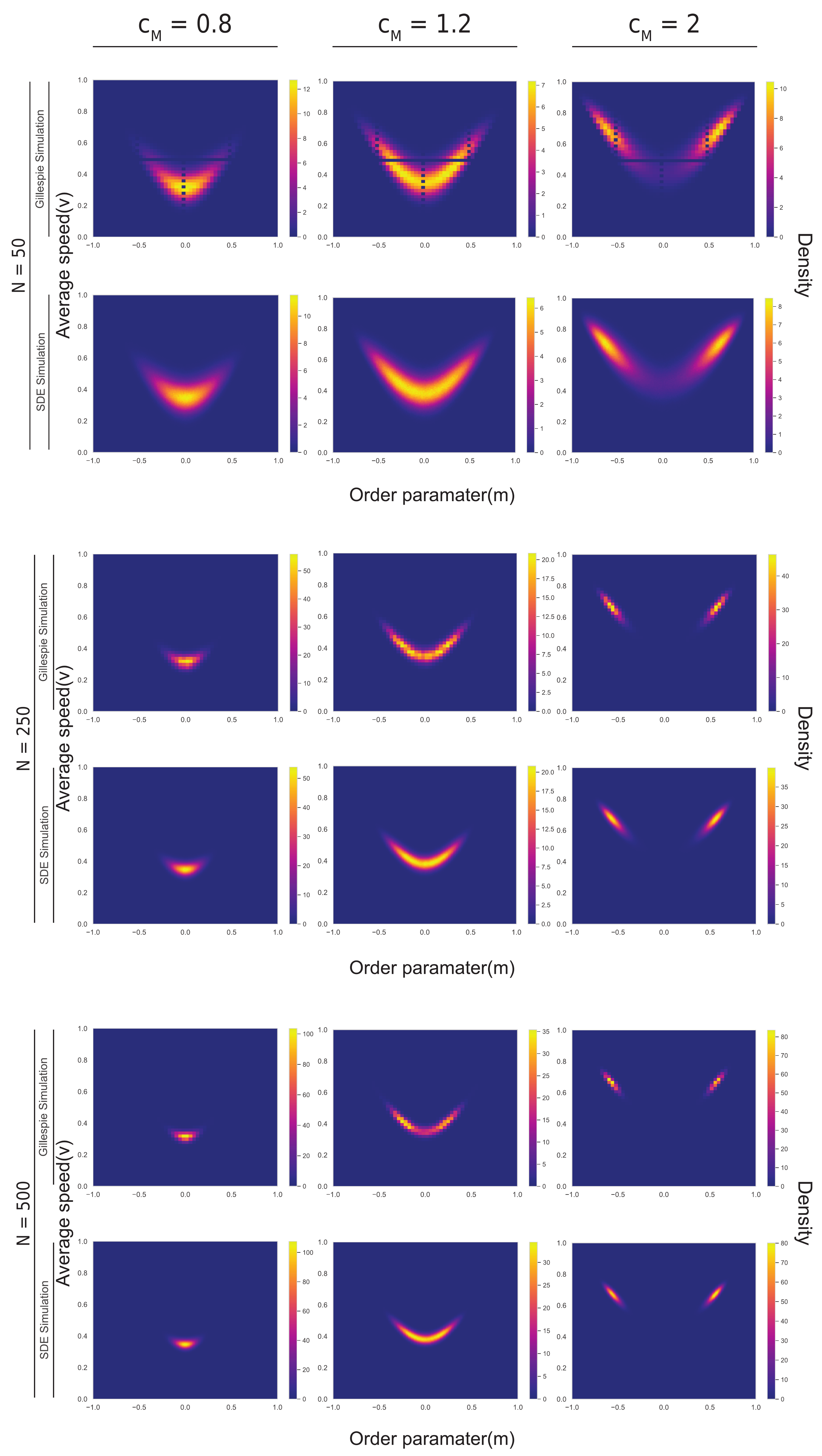}
    \caption{\textbf{Stochastic dynamics of the model.} The SDE approximation captures the actual dynamics of the system over a range of parameters.}
    \label{fig:noise}
\end{figure*}

\Cref{fig:noise} shows the joint distributions $m$ and $v$ of the model for different parameter regimes and different group sizes. For comparison, distributions are shown with both microscopic simulations as well as simulations of the SDE models. We see that, across parameter ranges as well as group sizes, the steady state distributions obtained from the SDE models closely matches those obtained from the stochastic simulations of the microscopic model.



\end{document}